# Five-channel frequency-division multiplexing using low-loss epsilon-near-zero metamaterial waveguide

Binbin Hong [1], Lei Sun [1], Wanlin Wang [1], Yanbing Qiu [1], Naixing Feng [1], Dong Su [2], Nutapong Somjit [3], Ian Robertson [3], and Guo Ping Wang [1*]

[1] *Institute of Microscale Optoelectronics, Shenzhen University, Shenzhen 518060, China*
[2] *The Tenth Research Institute, China Electronics Technology Group Corporation, Chengdu 610036, China*
[3] *School of Electronic and Electrical Engineering, University of Leeds, Leeds LS2 9JT, UK*
*Corresponding author (email: gpwang@szu.edu.cn)*

The rapidly growing global data usage has demanded more efficient ways to utilize the scarce electromagnetic spectrum resource. Recent research has focused on the development of efficient multiplexing techniques in the millimeter-wave band (1 – 10 mm, or 30 – 300 GHz) due to the promise of large available bandwidth for future wireless networks. Frequency-division multiplexing is still one of the most commonly-used techniques to maximize the transmission capacity of a wireless network. Based on the frequency-selective tunnelling effect of the low-loss epsilon-near-zero metamaterial waveguide, we numerically and experimentally demonstrate five-channel frequency-division multiplexing and demultiplexing in the millimeter-wave range. We show that this device architecture offers great flexibility to manipulate the filter Q-factors and the transmission spectra of different channels, by changing of the epsilon-near-zero metamaterial waveguide topology and by adding a standard waveguide between two epsilon-near-zero channels. This strategy of frequency-division multiplexing may pave a way for efficiently allocating the spectrum for future communication networks.

## 1 Introduction

The explosion of mobile data traffic has prompted the need for high data throughput to sustain high-bandwidth communication, but the electromagnetic spectrum has become extremely crowded due to the large and fast-growing number of new applications and end-users [1, 2]. Mobile networks can enhance the capacity of data throughput by increasing the number of non-interfering channels, stimulating the development of new multiplexing techniques. Various multiplexing techniques, including space, code, time, polarization, orbital angular momentum, and frequency, have been employed in different communication systems [3-6]. Among these techniques, frequency-division multiplexing (FDM) is one of the most commonly used solutions in modern wireless communication networks [7]. In the millimeter-wave range, where the spectrum has been split into a number of sub-bands to meet the diverse needs of satellite communication [8], 5G and 6G wireless communication [9] astronomical observation [10], security [11], etc., FDM is an even more important technique.

Previously, various approaches have been used to design multiplexers or demultiplexers (MUX/DEMUX), such as printed transmission lines, [12] metallic waveguides [13], surface or bulk acoustic resonators [14], photonic crystals [15], and dielectric ring or disc resonators [16, 17]. Normally, high Q resonators are needed in a MUX or DEMUX for selecting the target carrier frequencies, and these are very sensitive to the fabrication tolerances. Besides, if the operating frequencies of different channels are needed to be broadband, multistage cascaded resonators may be required, which increases the difficulties in design and fabrication. Here, we describe a device structure that exploits the tunnelling effect of an epsilon-near-zero (ENZ) metamaterial waveguide to act as the frequency-selective element in a MUX or DEMUX at Ka-band, and supports a travelling wave rather than a standing wave in the passband [18]. The Q-factor and bandwidth of each channel in the proposed device is highly sensitive to the topology of the ENZ waveguide, which offers the flexibility to manipulate the performance of the device.

The ENZ waveguide is a special kind of effective medium with near-zero real permittivity. In nature, the ENZ behavior can be found in noble metals [19, 20], plasmas [21], doped semiconductors [22], polar dielectrics [23], transparent conducting oxides [24, 25], etc. To effectively manipulate the electromagnetic wave, various artificial microstructures have been realized as the ENZ media, such as metal-coated waveguides near the cut-off frequency [18, 26-28], metallic nanowire arrays in the dielectric medium [29-31], metal-dielectric multilayer structures [32, 33], and Dirac-like cone-based metamaterial [34, 35]. Especially, several state-of-the-art experimental realizations have been successfully demonstrated in the microwave [18, 28] and optical [27, 29, 31-34] regions. Among the above effective ENZ media, the ENZ waveguide possesses relatively low insertion loss, making it a desirable medium for guiding the electromagnetic wave. The ENZ waveguide operates near the cut-off frequency of the metal waveguide, which breaks the rule-of-thumb for a standard hollow metallic rectangular waveguide (HMRW) normally operating between 125% and 189% of the cut-off frequency of the fundamental $TE_{10}$ mode [36]. The ENZ waveguide displays near-zero effective permittivity ($\varepsilon_{eff}$) and refractive index ($n_{eff}$) and has extremely large phase velocity and consequently large guided wavelength. Since the phase velocity of the ENZ waveguide approaches infinity, any media between the input and output faces of the ENZ waveguide can be physically far apart and yet but electrically near, and the ENZ waveguide will appear to be electrically small compared with the wavelength. Under this condition, the electromagnetic wave tends to tunnel through the ENZ waveguide following the Fermat's principle [37]. Recently, the ENZ waveguide has attracted much attention due to its exotic properties and has been experimentally demonstrated in both microwave [18] and optical frequency ranges [27]. Many functional devices have been reported based upon the ENZ waveguide, such as filters [38-41], sensors [42-45], waveguide transitions [46], antennas [47, 48], and modulator [49]. However, so far, little work has been reported on frequency-division multiplexing based on the ENZ waveguide.

This paper describes the experimental demonstration of a five-channel multiplexer/demultiplexer in the millimeter-wave range, based on the low-loss ENZ waveguide exploiting the tunnelling effect. By changing the width of the ENZ waveguide, the operating band of each channel can easily be adjusted, and by changing the height and length of the ENZ waveguide, the Q-factors can be flexibly manipulated. Additionally, by adding a standard hollow metallic rectangular waveguide (HMRW) between two ENZ channels, one can not only improve the Q-factor but also suppress the high-order transmission windows. The proposed device supports both MUX and DEMUX capabilities as their transmission properties are almost identical. The proposed structure can offer a solution for frequency division multiplexing in the millimeter-wave range that does not employ cavity resonators or similar structures that require precision machining.

## 2 Methods

### 2.1 Fabrication

The device under consideration (see Figure 1(a), inset) is made of T2 electrolytic-tough-pitch (ETP) copper whose electrical conductivity is larger than $5.6 \times 10^7$ S/m. The grooves and holes of the proposed structure are milled or drilled by using the high-speed 3-axis computer numerical control (CNC) milling machine JDHGT600T from Beijing Jingdiao®, which has a milling resolution of better than 5 μm. The structure is split into two separate parts mainly by the common bottom plane of the standard rectangular waveguides and the ENZ waveguides. The two parts are tightly combined with the aid of alignment holes and grooves. Sidewalls between each channel, acting as electrically shorted terminations, are used to prevent signal leakage and interference between channels. All geometrical parameters of the proposed structure are given in Figure 1(b) and Table 1.

### 2.2 Simulation

The CST Microwave Studio time-domain solver has been used to simulate the transmission behaviors of the ENZ multiplexer or demultiplexer. The air channels are surrounded by lossy metal whose electrical conductivity is set as $5.6 \times 10^7$ S/m. The dispersive surface impedance of the lossy metal is included in the simulation using an internal one-dimensional surface impedance model which takes the skin effect into account. The metal surfaces are set to be smooth, so the surface roughness is not considered for the purpose of simplicity. A waveguide port, which can also be regarded as a perfect impedance matched load, is used to excite and detect the signal at each port.

### 2.3 Measurement

The device was characterized using a Keysight N5225A two-port PNA Vector Network Analyzer. Each port of the device is surrounded by standard WR28 rectangular waveguide flange apertures. During the measurement, the two ports under test are fed by coaxial cables, which are connected with the PNA, using coaxial-to-rectangular waveguide transitions as interconnects, and meanwhile, the other ports remain open. The different load situations of the ports between simulation and measurement contribute to the slight differences in the results, but the impact is limited as the crosstalk between channels is insignificant. The voltage standing wave ratio (VSWR) of the coaxial-to-rectangular waveguide transition is below 1.2 across the Ka-band. A Thru-Reflect-Line (TRL) calibration was performed using a standard 2.4 mm mechanical calibration kit to place the reference planes to the ends of the coaxial cables. The coaxial-to-rectangular waveguide transitions have not been de-embedded from the measurement, but the transitions are designed to have low insertion and return losses and hence only have a limited impact on the measurement results.

## 3 Demultiplexer and Multiplexer

### 3.1 Theoretical principle

The device under consideration is illustrated in Figure 1(a). The channel 1 ($C_1$) carries the multiplexed broadband signal, while the channels 2 to 6 ($C_2 – C_6$) carry the individual signals with different and narrow frequency-selected bands. $C_1$ interconnects with each of the other channels through two identical sections of narrow ENZ waveguide, as well as a section of standard rectangular waveguide between the two ENZ waveguides that acts as an impedance matching element. Here, the ENZ waveguide is a narrow air-filled metallic rectangular waveguide with reduced width and height, compared with the standard HMRW, working near the cut-off frequency of the fundamental $TE_{10}$ mode. Apart from the ENZ waveguides, all other waveguide sections are standard WR28 HMRW, whose width and height are 7.112 mm and 3.556 mm, respectively. The standard WR28 hollow metallic rectangular waveguide operates across Ka-band, from 25.6 to 40 GHz. When the device is used as a demultiplexer, a broadband signal is input via the $C_1$ and selectively routed to the other channels depending on the frequency. *Vice versa*, when acting as a multiplexer, signals of different frequencies from different channels ($C_2 – C_6$) are combined into $C_1$ for simultaneous transmission as one broadband signal.

The multiplexing principle originates from the tunnelling phenomenon of the ENZ waveguides, which allows the electromagnetic wave to tunnel through the narrow ENZ waveguide in spite of the large impedance mismatch at its entrance and exit faces [37]. This phenomenon occurs near the cut-off frequency of the $TE_{10}$ mode of the ENZ waveguide. Considering an ideal case where the metal wall is set as a perfect electric conductor (PEC) and has no surface roughness, no Ohmic loss occurs in the HMRW. The propagation constant for the fundamental transverse-electric $TE_{10}$ mode of an air-filled hollow metallic rectangular waveguide can be expressed as

$$\beta = \sqrt{\left(\frac{2\pi f}{c}\right)^2 - \left(\frac{\pi}{w}\right)^2} \qquad (1)$$

Here, $w$ is the waveguide $H$-plane width, $c$ is the speed of light, and $f$ is the operating frequency. The effective refractive index of the $TE_{10}$ mode is $n_{eff} = \beta/k_0$, where $k_0$ is the free-space wavenumber. The effective permittivity of the fundamental $TE_{10}$ mode in the HMRW is given by [50]:

$$\varepsilon_{eff} = \varepsilon_0 \left(1 - \frac{c^2}{4f^2 w^2}\right) \qquad (2)$$

This shows that it is an inherently frequency dispersive medium. The effective permeability, $\mu_{eff}$, for the $TE_{10}$ mode remains $\mu_0$ which is the vacuum permeability. At the cut-off frequency of this dominant $TE_{10}$ mode, namely $f_c = c/2w$, it is found that $\beta = n_{eff} = \varepsilon_{eff} = 0$, which leads to the infinite phase velocity of the wave propagation in the ENZ waveguide.

In practice, the cut-off frequency of the ENZ waveguide plays a prominent role in determining the most efficient tunnelling behavior. Other factors, including the ohmic loss due to surface conductivity, the scattering loss due to the surface roughness, and especially the input and output impedances at the entrance and exit faces of the ENZ waveguide, shift the most efficient tunnelling point slightly away from the cut-off frequency of the ENZ waveguide. It can either be above or below the cut-off frequency, depending on the input and output impedances [51].

The cut-off frequencies of the ENZ waveguides for all five channels in the proposed MUX and DEMUX are firstly calculated based on Eq. (1) by setting $\beta = 0$, and the chosen values are illustrated in Figure 2. For each channel, the widths, heights, and lengths of the two narrow ENZ waveguides are all the same. The cut-off frequencies of all ENZ waveguides, denoted as $F_a$, $F_b$, $F_c$, $F_d$, and $F_e$, are designed to be within the single-mode operating frequency range of the standard WR28 HMRW, which is above the cut-off frequency of its fundamental $TE_{10}$ mode and below the cut-off frequency of its second-order $TE_{20}$ mode.

### 3.2 Simulation and Measurement Results

When acting as a DEMUX, the signal is inputted via the $C_1$ main channel and detected at the $C_2 – C_6$ subchannels. The simulated and measured transmission coefficients of all channels are presented in Figure 3. The frequencies of interest span across the Ka-band, ranging from 26.5 to 40 GHz. In both simulation and measurement results, five channels can be observed from the transmission spectra. The simulated and measured results of center frequencies ($f_o$), -3 dB cut-off frequencies ($f_L$ and $f_H$), bandwidths ($\Delta f$), and Q-factors ($Q$) are listed in Table 2. Here, $\Delta f = f_H - f_L$, $f_o = \sqrt{f_H f_L}$, and $Q = f_0/\Delta f$. The difference in center frequencies between the simulation and the measurement results for all channels is $0.402 \pm 0.104$ GHz. The simulated and measured Q-factors range from 20.65 to 39.31 and from 18.20 to 35.54, respectively.

From the zoomed-in transmission spectra shown in Figures 3(c) and 3(d), the measured magnitudes of all channels are slightly lower than that of the simulated results. This is mainly attributed to the additional WR28 coaxial-to-rectangular waveguide transitions which haven't been de-embedded in the calibration due to the lack of a standard calibration kit for the WR28 HMRW. The other factor that contributes to the difference is the presence of surface roughness on the waveguide walls, which causes additional Ohmic loss compared to the smooth walls in the simulation. The measured central operating frequencies of each channel are slightly lower than that of the simulated results, which is primarily caused by the fabrication imperfections, as well as the different loads of ports between the simulation and measurement. The loads in the simulation are all matched, while they are open for ports not under testing in the measurement. However, its impact is limited as the isolation between different channels is better than 15 dB, which will be further discussed later. Nonetheless, from the perspectives of the operating bands, the magnitudes of the transmitted signals, and the shapes of the transmission spectra, the measurement results are still highly consistent with the simulation results, demonstrating that the fabricated device can efficiently work as a five-channel DEMUX.

The numerically simulated normalized electric fields propagating in the DEMUX at five representative frequencies are presented in Figure 4. The five representative frequencies correspond to the lowest insertion loss points of the five channels presented in Figure 3(c). The broadband signal spanning over 26.5 to 40 GHz is input via the $C_1$ channel. It can be seen that the signals with different frequencies are successfully routed to different channels based on the selection mechanism of the DEMUX. The electric fields in the unused channels are well suppressed by the ENZ waveguides. As the ENZ waveguides are much thinner in height and slightly narrower in width when compared with the standard WR28 HMRW, the intensity of the electric field in the ENZ waveguide is significantly enhanced. Besides, according to the simulation, the phases of the propagating $TE_{10}$ mode along the ENZ waveguides are essentially constant, which is in agreement with the theoretical analysis discussed in Section 3.1.

The channel-to-channel cross talk in the proposed ENZ DEMUX has been experimentally studied, as shown in Figure 5. The cross talk between neighbor frequency channels is lower than -15 dB, and that of non-neighbor channels is lower than -30 dB, demonstrating that good isolation has been achieved in the proposed ENZ DEMUX. Reducing the cross talk is one of the main targets of designing a MUX or DEMUX, but it is also challenging, especially for multi-channel devices. To further reduce the cross talk in the proposed device, it is needed to improve the q-factors of different channels and meanwhile prevent high-order transmission windows from rising within the operating band, which will be discussed in Section 4.

Acting as a MUX, the signals with different frequencies from $C_2$-$C_6$ are combined in $C_1$. The measured transmission spectra of all channels are presented in Figure 6. Comparing Figure 5 with Figure 3(b), the MUX's spectra are almost identical to the DEMUX's spectra, indicating the proposed device can work both as a DEMUX and as a MUX. The performances of the MUX are similar to that of the DEMUX listed in Table 2.

## 4 Manipulation of the Q-factor

Q-factor is the key factor to determine the operating bandwidth, cut-off rate, and cross talk of the proposed DEMUX and MUX. However, improving the Q-factor is not an isolated issue. Simultaneously, one has to consider the insertion loss of different frequency channels, the overlap of high-order transmission windows with other channels, as well as the fabrication difficulties. The Q-factors of the proposed device are mainly determined by the topology of the ENZ channel. The physical insight behind this is the impedance matching condition between the input and output ports along the cascaded structure, which has been discussed in [37] We also find that by adding a standard HMRW between two ENZ channels, one can not only improve the Q-factor but also suppress the high-order transmission window.

Changing the parameters of the ENZ channel, including the height and length, is the traditional way to manipulate the Q-factor of the ENZ-based frequency filtering structures [18]. Figure 7(a) shows a three-channel DEMUX and the transmission spectra are presented in Figures 7(b) and 7(c), for different height or length, respectively, of the ENZ waveguide. Here, the red, green, blue curves correspond to the $S_{21}$, $S_{31}$, and $S_{41}$, respectively. In both Figures 7(b) and 7(c), the widths of the ENZ waveguides for channels 2, 3, and 4 are 5.2 mm, 4.3 mm, and 3.7 mm. In Figure 7(b), the solid, dotted, and dash-dotted curves represent the heights of the ENZ waveguides of 0.5 mm, 0.3 mm, and 0.1 mm. In Figure 7(c), the solid, dotted, and dash-dotted curves represent lengths of the ENZ waveguides of 1.5 mm, 2.5 mm, and 3.5 mm, respectively. The Q-factor rises with the decrease of height and the increase of length, while it also involves an increasing insertion loss as a penalty. Besides, with the increasing length of the ENZ waveguide, the transmission windows of high-order Fabry-Perot modes get redshifted, which probably will overlap with the other channels of the DEMUX causing the cross-talk problem. In addition, the decreasing height of the ENZ waveguide would concentrate the current density significantly, introducing large Ohmic loss. With only one section of the ENZ waveguide, the degree of freedom to further optimize the transmission spectra is limited.

Introducing an additional standard HMRW in-between two sections of the ENZ waveguide allows more freedom to deal with the above problems, as shown in Figure 7(d). The additional standard HMRW is not a frequency selective resonator, however, unlike in most conventional filters or multiplexers. At the operating ENZ frequency, the wave can tunnel through the ENZ channels at the input and output faces of the additional HMRW, so no standing wave is formed. However, at rejection frequencies, the additional HMRW further enlarges the impedance mismatch between the narrow waveguide and the HMRW, suppressing the signal transmission. Besides, it also breaks the resonance condition of the high-order Fabry-Perot modes, and hence makes these modes shift to blue rather than red.

Figures 7(e) and 7(f) present the transmission coefficients before and after the inclusion of the additional HMRW. Here, the heights of channels 2, 3, and 4 are 0.52 mm, 0.48 mm, and 0.44 mm, and their lengths are 1.8 mm, 2.0 mm, and 2.2 mm, and their widths are 5.2 mm, 4.3 mm, and 3.7 mm, respectively. In Figure 7(e), the high-order Fabry-Perot mode of the $P_4$ channel overlaps with the ENZ mode of the $P_2$ channel at or near 39 GHz, causing the transmission reduction of the $P_2$ channel and the increase of the cross talk between the $P_2$ and $P_4$ channels. Besides, the $P_4$ channel also transmits a significant amount of power of the frequencies that overlapped with the $P_3$ channel, which reduces the transmission of the $P_3$ channel and also introduces significant cross talk between the $P_3$ and $P_4$ channels. In contrast, when the additional HMRW is placed in the middle of the ENZ waveguide, as shown in Figure 7(f), the transmission and Q-factors of all channels are improved, and the cross talk between different channels is greatly reduced compared with Figure 7(e). Here, the lengths of the additional HMRWs are 3.3 mm, 2.9 mm, 4.0 mm for channels 2, 3, and 4, respectively. Therefore, introducing an additional HMRW in-between the ENZ waveguides is an efficient way to improve the Q-factor, suppress the high-order modes, reduce the cross talk, and yet maintain low insertion losses for all channels. More interestingly, it is possible to individually manipulate the Q-factors of each channel to select the bandwidth of the target carrier frequencies passing through it.

The fabricated five-channel structure, illustrated in Figure 1(a), shows an optimized design considering the number of channels, Q-factor of each channel, cross talk, high-order mode competition, and fabrication tolerance. It is possible to further enhance the Q-factor and reduce the cross talk, but this may involve trade-offs such as the cost, reducing the number of channels over the desired Ka-band, or increasing fabrication difficulties using the currently employed CNC milling technique. Moreover, placing more subchannels is of great interest in practical devices. However, there are limitations on the overall number of the individual ports that can be attached, e.g. the total available single-mode operation bandwidth of the main channel, and the bandwidths of the subchannels with acceptable insertion losses. One has to balance the desired Q-factors with the acceptable insertion losses of the subchannels when determining the geometrical dimension of the ENZ-based interconnecting parts. The design is scalable and can be adapted at other frequencies, such as terahertz or optical bands, since the ENZ waveguide operating at the optical frequencies has also been demonstrated [26].

# 5 Conclusions

In summary, we have experimentally demonstrated a five-channel frequency-division MUX/DEMUX based on low-loss epsilon-near-zero metamaterial waveguide in the millimeter-wave frequency range of 26.5 to 40 GHz. Unlike other MUX/DEMUX structures using resonators as the frequency-selective element, our approach exploits the tunnelling effect of the ENZ waveguide. This supports a travelling wave rather than a standing wave in the passband, offering a novel design methodology. Varying the width of the ENZ waveguide, which changes the cut-off frequency, provides a straightforward method to adjust the operating bands of the different channels. Moreover, we have also demonstrated that our device structure offers great flexibility to manipulate the Q-factors and transmission spectra of different channels, by changing of the height and length of the ENZ waveguide and by adding a standard waveguide between two ENZ channels. The measured insertion losses of all channels are lower than 1.8 dB and the Q-factors are measured to be within the range from 18.20 to 35.54. The crosstalk between neighboring frequency channels is lower than -15 dB, and that of non-adjacent channels is lower than -30 dB. The proposed strategy of frequency-division multiplexing may pave a way for efficiently allocating the spectrum for future wireless communication networks.

*The authors thank Lei Ge and Shuai Gao for their help during the measurement. This project was partially supported by the National Natural Science Foundation of China (11734012, 62105213, 12074267, 516022053, 12174265), the Young Innovative Talents Project of Universities in Guangdong Province (2019KQNCX123), the Guangdong Basic and Applied Basic Research Fund (2020A1515111037), the Science and Technology Project of Guangdong (2020B010190001), the Guangdong Natural Science Foundation (2020A1515010467), the Shenzhen Fundamental Research Program (202008141136250003), and the Open Fund of State Key Laboratory of Applied Optics (SKLAO2020001A06).*

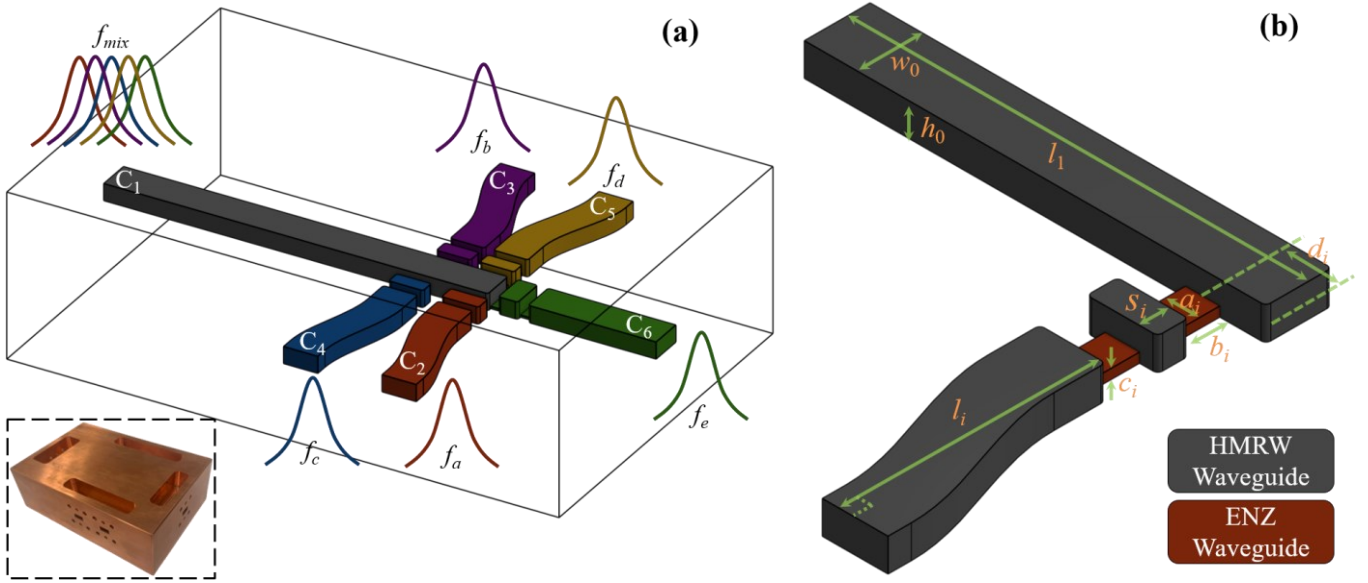

**Figure 1** (Color online) The proposed MUX or DEMUX. (a) The full five-channel structure. The colored regions are air channels surrounded by copper cladding (transparent). Channel 1 ($C_1$) conveys a broadband signal which is divided into (MUX) or combined from (DEMUX) Channels 2-6 ($C_2$-$C_6$). The narrow waveguides interconnecting the main channel and the subchannels are ENZ waveguides with different cut-off frequencies. A standard rectangular waveguide is placed between two ENZ waveguides in each channel for improving the Q-factor and suppressing the high-order transmission modes. The inset shows the fabricated sample. (b) Single channel of the MUX or DEMUX with key geometric parameters indicated. The gray and red regions are air channels surrounded by copper cladding which is not shown in the figure. The gray regions are standard HMRW whose width ($w_0$) and height ($h_0$) are 7.112 mm and 3.556 mm, respectively. The red regions are ENZ waveguides interconnecting the main channel and the subchannels. An additional HMRW is placed between the two ENZ waveguides and its length is $S_i$. Here, $i$ refers to the identifier of subchannels, and $i \in \{2, 3, 4, 5, 6\}$. Bends are applied to the inner right corners of the multiplexer for fabrication purpose. The bend radius is 0.5 mm. The detailed geometric parameters for the full five-channel MUX/DEMUX are shown in Table 1.

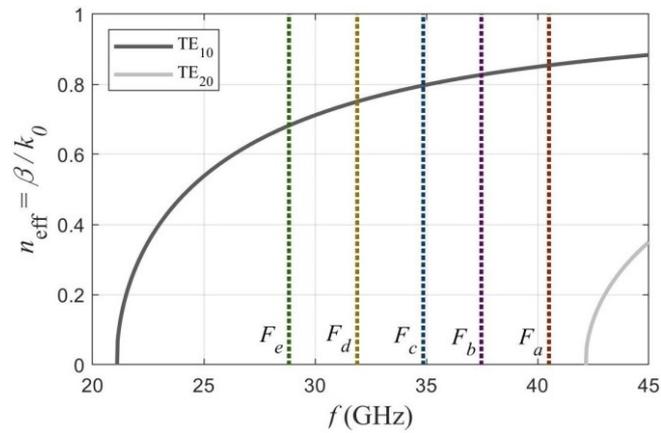

**Figure 2** (Color online) Dispersion curves and cut-off frequencies. The solid black and gray curves represent the dispersion curves corresponding to the $TE_{10}$ and $TE_{20}$ modes, respectively, of a standard WR28 hollow metallic rectangular waveguide. The vertical dashed lines indicate the cut-off frequencies of the five ENZ waveguides. $F_a$, $F_b$, $F_c$, $F_d$, and $F_e$ correspond to the ENZ waveguide widths of 3.7 mm, 4.0 mm, 4.3 mm, 4.7 mm, and 5.2 mm, respectively.

**Table 1: The detailed geometric parameters of the full five-channel MUX/DEMUX[*].**

|       | $h_0$ |       | $w_0$ |       |       | $l_0$ |
|-------|-------|-------|-------|-------|-------|-------|
| $C_1$ | 3.556 |       | 7.112 |       |       | 80    |
|       | $a_i$ | $b_i$ | $c_i$ | $d_i$ | $s_i$ | $l_i$ |
| $C_2$ | 3.7   | 1.8   | 0.52  | 1     | 3.3   | 28.1  |
| $C_3$ | 4     | 1.9   | 0.52  | 10.4  | 3.1   | 28.1  |
| $C_4$ | 4.3   | 2     | 0.48  | 12.1  | 2.9   | 28.1  |
| $C_5$ | 4.7   | 2.1   | 0.5   | 1.1   | 2.9   | 27.9  |
| $C_6$ | 5.2   | 2.2   | 0.44  | 0     | 4     | 26.6  |

[*] The unit for this table is mm.

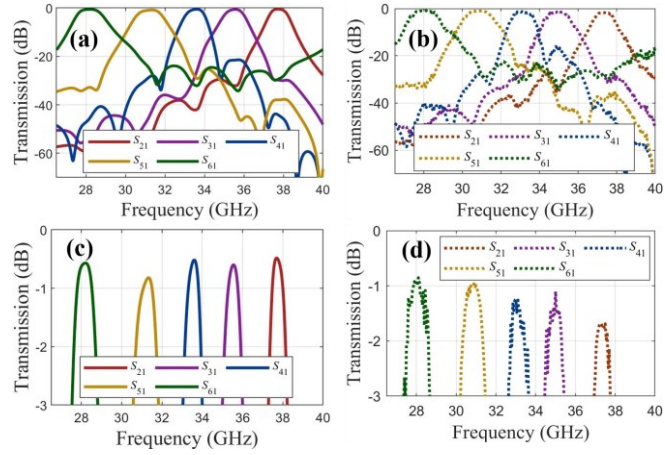

**Figure 3** (Color online) The simulated and measured transmission coefficients of the demultiplexer. (a) Simulated transmission coefficients. (b) Measured transmission coefficients. (c) Zoomed simulated transmission coefficients. (d) Zoomed measured transmission coefficients.

**Table 2: The Simulated and Measured Transmission Properties of Different Subchannels.**

| Transmission Coefficients | | Center Frequency $f_0$ (GHz) | Cut-off Frequencies (GHz) | | Bandwidth $\Delta f$ (GHz) | Quality Factor $Q$ |
|---|---|---|---|---|---|---|
| | | | $f_L$ | $f_H$ | | |
| Sim. | $S_{21}$ | 37.74 | 37.26 | 38.22 | 0.96 | 39.31 |
|      | $S_{31}$ | 35.52 | 34.98 | 36.06 | 1.08 | 32.89 |
|      | $S_{41}$ | 33.53 | 33.03 | 34.03 | 1.00 | 33.53 |
|      | $S_{51}$ | 31.23 | 30.55 | 31.93 | 1.38 | 22.63 |
|      | $S_{61}$ | 28.29 | 27.61 | 28.98 | 1.37 | 20.65 |
| Exp. | $S_{21}$ | 37.32 | 36.80 | 37.85 | 1.05 | 35.54 |
|      | $S_{31}$ | 34.97 | 34.40 | 35.55 | 1.15 | 30.41 |
|      | $S_{41}$ | 33.15 | 32.58 | 33.72 | 1.14 | 29.07 |
|      | $S_{51}$ | 30.83 | 30.12 | 31.56 | 1.44 | 21.41 |
|      | $S_{61}$ | 28.03 | 27.27 | 28.81 | 1.54 | 18.20 |

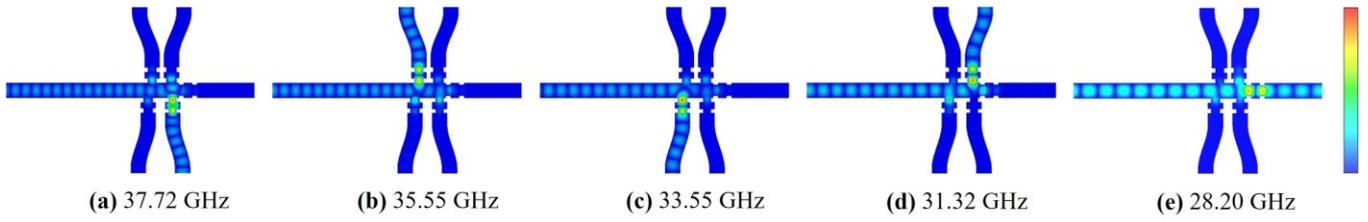

(a) 37.72 GHz  (b) 35.55 GHz  (c) 33.55 GHz  (d) 31.32 GHz  (e) 28.20 GHz

**Figure 4** (Color online) Normalized electric field on a log scale at several representative frequencies. The field exponentially decreases from red to blue.

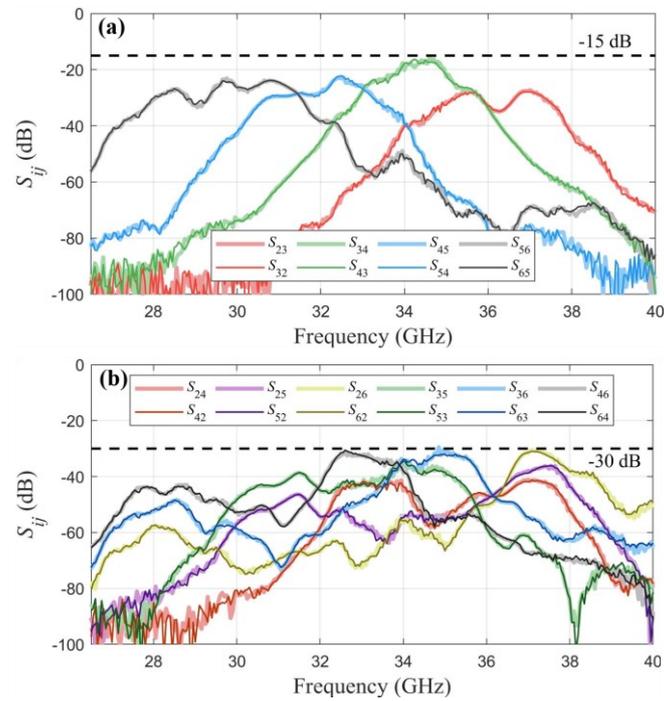

**Figure 5** (Color online) Measured channel-to-channel cross talk. Here, $i$ and $j$ denote the channel number, and $i \neq j$. (a) Cross talk between neighbor frequency channels. (b) Cross talk between non-neighbor frequency channels.

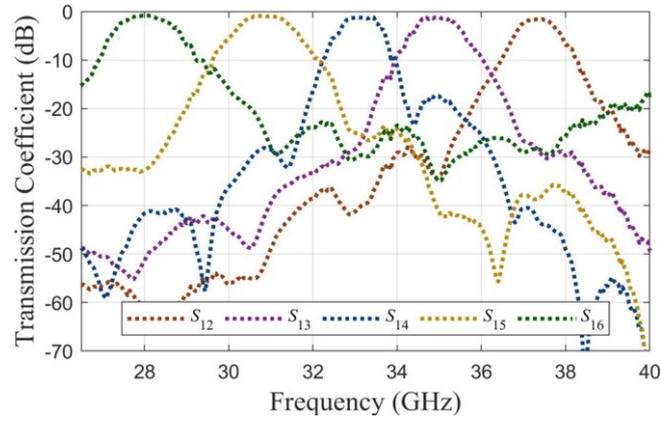

**Figure 6** (Color online) Measured transmission coefficients of the multiplexer.

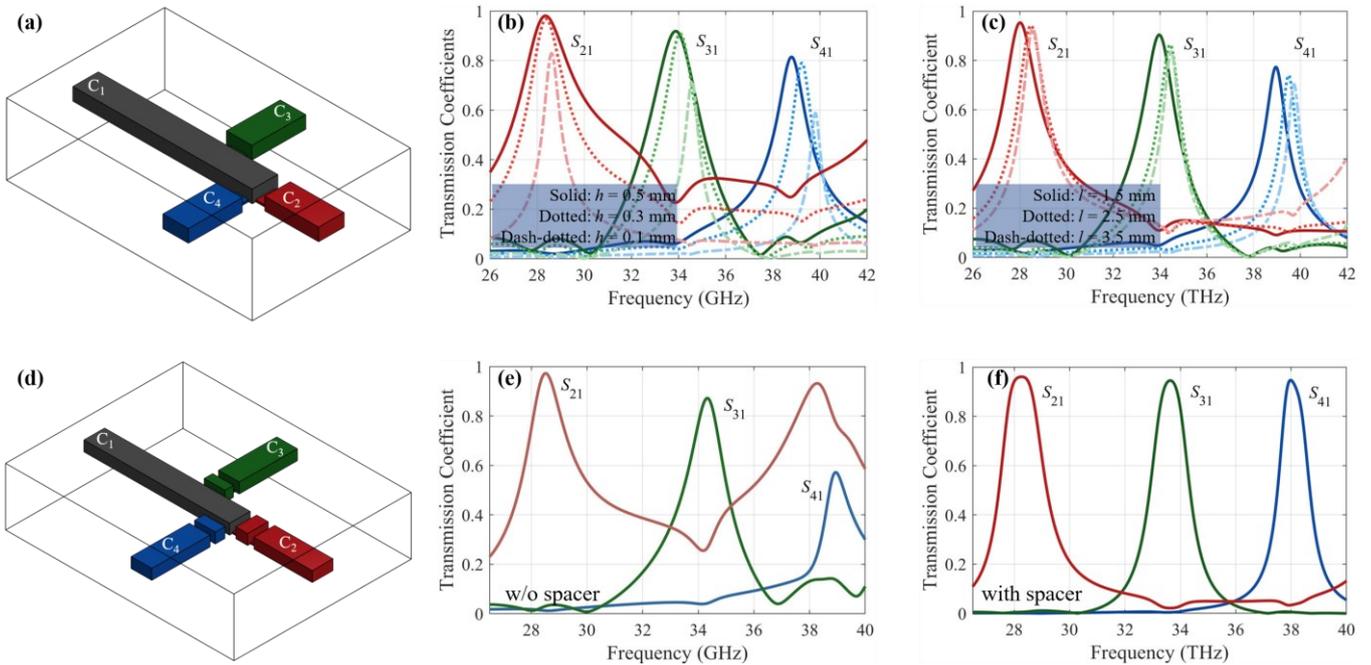

**Figure 7** (Color online) Q-factor manipulation. (a) Schematic of the three-channel DEMUX using a single and identical ENZ waveguide as the frequency-selective element for each channel. The input or output ports of the black, red, green, blue channels are denoted as ports 1, 2, 3, and 4, respectively, which also applies to Figure 7(d). (b) Transmission coefficients with different heights of the ENZ waveguides. Here, the solid, dotted, and dash-dotted curves represent the heights of the ENZ waveguides ($h$) of 0.5 mm, 0.3 mm, and 0.1 mm, respectively. (c) Transmission coefficients with different lengths of the ENZ waveguides. Here, the solid, dotted, and dash-dotted curves represent the lengths of the ENZ waveguides ($l$) of 1.5 mm, 2.5 mm, and 3.5 mm, respectively. (d) Schematic of the three-channel DEMUX using two ENZ waveguides and an additional HMRW waveguide as the frequency-selective element. (e) Transmission coefficients without (w/o) the additional HMRW waveguide. (f) Transmission coefficients with the additional HMRW waveguide.